# To Detect Irregular Trade Behaviors In Stock Market By Using Graph Based Ranking Methods

Loc Tran[1], Linh Tran[2]

[1]*John von Neumann Institute, VNU-HCM; tran0398@umn.edu*

[2]*Thu Dau Mot University, Binh Duong Province, Vietnam; linhtran.cntt@tdmu.edu.vn*

Abstract: To detect the irregular trade behaviors in the stock market is the important problem in machine learning field. These irregular trade behaviors are obviously illegal. To detect these irregular trade behaviors in the stock market, data scientists normally employ the supervised learning techniques. In this paper, we employ the three graph Laplacian based semi-supervised ranking methods to solve the irregular trade behavior detection problem. Experimental results show that that the un-normalized and symmetric normalized graph Laplacian based semi-supervised ranking methods outperform the random walk Laplacian based semi-supervised ranking method.

**Introduction**

In the stock market, there are some investors who try to get benefits from the stock market by using irregular trade behaviors that affect the stock price. These irregular trade behaviors include pump-and-dump and spoof trading [1]. These trade behaviors are illegal. Moreover, the control of these irregular trade behaviors is very difficult due to the large amount of transaction data. In order to detect these irregular trade behaviors, data scientists employ a lot of supervised learning techniques such as Neural Network and Support Vector Machine [2], to name a few.

However, to the best of our knowledge, the graph based semi-supervised ranking techniques have not been applied to this irregular trade behaviors detection problem.

Hence, in the other words, in this paper, we will try predict the new members of a partially known set of transactions involved in irregular trade behaviors in the transaction-transaction network (i.e. graph). In this problem, we are given a core set of transactions (i.e. the queries) involved in irregular trade behaviors. However, the financial experts do not know whether this core set is complete or not. Our objective is to find more potential members of the set of transactions involved in irregular trade behaviors by ranking transactions in transaction-transaction network. Then the transactions with highest ranks (i.e. probability of membership in the partially known set of transactions involved in irregular trade behaviors) will then be selected and checked by financial experts to see if the extended transactions in fact belong to the core set of transactions involved in irregular trade behaviors.

Most network-based algorithms have been proposed to include topological properties of transaction-transaction networks in understanding irregular trade behaviors (i.e. local methods). These algorithms mainly utilize the idea that the transactions associated with irregular trade behaviors have a higher possibility of being connected in the transaction-transaction networks. However, a significant challenge for these applications is the partial and noisy nature of the transaction-transaction networks. Missing edges and false positives affect the accurateness of

"local methods" based on local information such as edge weights and shortest distances.

Few "global methods" based on simulation of information flow in the network (e.g., random walks [3,4] or network propagation [5,6]) avoid this problem by considering numerous different paths and the whole topology of transaction-transaction networks.

To solve this irregular trade behaviors detection problem, we can simulate a random walker that starts from a set of nodes (i.e. the queries or the set of transactions involved in irregular trade behaviors) instead of a single node. Thus, given a set of transactions that is involved in irregular trade behaviors as the start set, the random walk on graphs method (i.e. the random walk graph Laplacian based semi-supervised ranking method) ranks the remaining transactions in the transaction-transaction network with respect to their proximity to the queries. This ranking method [3,4] has also been employed by Google Company to exploit the global hyperlink structure of the Web and produce better rankings of search results [7]. Its idea [3,4,7] has also been employed in [8] to solve the protein function prediction problem. However, based on [8], the random walk on graphs method is not the best network-based ranking method. Moreover, the random walk on graphs method has not been applied to irregular trade behaviors detection problem.

Unlike the random walk on graphs method utilizing the random walk graph Laplacian, the network propagation method (i.e. the symmetric normalized graph Laplacian based semi-supervised ranking method) [5,6] employs the symmetric normalized graph Laplacian. However, similar to the random walk of graphs method, the network propagation method has not been applied to irregular trade behaviors detection problem.

Moreover, to the best of our knowledge, the un-normalized graph Laplacian based semi-supervised ranking method is considered the current state of the art network-based ranking method solving protein function prediction problem [9,10]. However, the un-normalized graph Laplacian based ranking method has not been applied to the irregular trade behaviors detection application.

Thus, in this paper, we will try to detect the transactions involved in the pump-and-dump trade behaviors (i.e. one specific irregular trade behaviors) by using un-normalized graph Laplacian based semi-supervised ranking method, the random walk graph Laplacian based semi-supervised ranking method, and the symmetric normalized graph Laplacian based semi-supervised ranking method.

We will organize the paper as follows: Section 2 will introduce random walk and symmetric normalized graph Laplacian based semi-supervised ranking algorithms in detail (iterative version). Section 3 will show how to derive the closed form solutions of the symmetric normalized and un-normalized graph Laplacian based semi-supervised ranking algorithms from regularization framework. In section 4, we will apply these three algorithms (i.e. the un-normalized, the random walk, and the symmetric normalized graph Laplacian based semi-supervised ranking algorithms) to the network constructed from the transaction datasets that are available from [11]. Section 5 will conclude this paper and the future direction of researches of this problem will be discussed.

## Algorithms

Given a set of samples $\{x_1, \ldots, x_l, x_{l+1}, \ldots, x_{l+u}\}$ where $n = l + u$ is the total number of samples in the network $W$. The way constructing $W$ will be discussed in Section 4.

Please note that $\{x_1, \ldots, x_l\}$ is the set of all ranked samples and $\{x_{l+1}, \ldots, x_{l+u}\}$ is the set of all un-ranked samples.

Let $Y \in R^n$, the initial ranking matrix for $n$ samples in the network, be defined as follows

$$Y_{ij} = \begin{cases} 1 \text{ if } x_i \text{ is the query} \\ 0 \text{ if } x_i \text{ is not the query and } 1 \leq i \leq l \\ 0 \text{ if } l+1 \leq i \leq n \end{cases}$$

Our objective is to predict the ranks of the un-ranked samples $x_{l+1}, \ldots, x_{l+u}$.

Let the matrix $F \in R^n$ be the estimated ranking matrix for the set of samples $\{x_1, \ldots, x_l, x_{l+1}, \ldots, x_{l+u}\}$.

We can achieve this objective by letting every nodes (i.e. samples) in the network iteratively propagates its ranking information to its adjacent nodes and this process is repeated until convergence. These three algorithms are based on three assumptions:

- local consistency: nearby samples likely have the same rank values
- global consistency: samples on the same structure (cluster or sub-manifolds) likely have the same rank values
- the network contains no self-loops

### Random walk graph Laplacian based semi-supervised ranking algorithm

In this section, we slightly change the original random walk graph Laplacian based semi-supervised ranking algorithm can be obtained from [8,12]. The outline of the new version of this algorithm is as follows

1. Construct $S_{rw} = D^{-1}W$ where $D = diag(d_1, d_2, \ldots, d_n)$ and $d_i = \sum_j W_{ij}$
2. Iterate until convergence
$F^{(t+1)} = \alpha S_{rw} F^{(t)} + (1-\alpha)Y$, where $\alpha$ is an arbitrary parameter belongs to $[0,1]$
3. Let $F^*$ be the limit of the sequence $\{F^{(t)}\}$.

Next, we look for the closed-form solution of the random walk graph Laplacian based semi-supervised ranking algorithm. In the other words, we need to show that

$$F^* = \lim_{t \to \infty} F^{(t)} = (1-\alpha)(I - \alpha S_{rw})^{-1}Y$$

Suppose $F^{(0)} = Y$, then

$$F^{(1)} = \alpha S_{rw} F^{(0)} + (1-\alpha)Y$$

$$= \alpha S_{rw} Y + (1-\alpha)Y$$

$$F^{(2)} = \alpha S_{rw} F^{(1)} + (1-\alpha)Y$$

$$= \alpha S_{rw}(\alpha S_{rw}Y + (1-\alpha)Y) + (1-\alpha)Y$$

$$= \alpha^2 S_{rw}^2 Y + (1-\alpha)\alpha S_{rw}Y + (1-\alpha)Y$$

$$F^{(3)} = \alpha S_{rw} F^{(2)} + (1-\alpha)Y$$

$$= \alpha S_{rw}(\alpha^2 S_{rw}^2 Y + (1-\alpha)\alpha S_{rw}Y + (1-\alpha)Y) + (1-\alpha)Y$$

$$= \alpha^3 S_{rw}^3 Y + (1-\alpha)\alpha^2 S_{rw}^2 Y + (1-\alpha)\alpha S_{rw}Y + (1-\alpha)Y$$

…

Thus, by induction,

$$F^{(t)} = \alpha^t S_{rw}^t Y + (1-\alpha)\sum_{i=0}^{t-1}(\alpha S_{rw})^i Y$$

Since $S_{rw}$ is the stochastic matrix, its eigenvalues are in [-1,1]. Moreover, since $0 \le \alpha \le 1$, thus

$$\lim_{t \to \infty} \alpha^t S_{rw}^t = 0$$

$$\lim_{t \to \infty} \sum_{i=0}^{t-1}(\alpha S_{rw})^i = (I - \alpha S_{rw})^{-1}$$

Therefore,

$$F^* = \lim_{t \to \infty} F^{(t)} = (1-\alpha)(I - \alpha S_{rw})^{-1} Y$$

Now, from the above formula, we can compute $F^*$ directly.

### Symmetric normalized graph Laplacian based semi-supervised ranking algorithm

Next, we will give the brief overview of the original normalized graph Laplacian based semi-supervised ranking algorithm can be obtained from [6,8]. The outline of this algorithm is as follows

1. Construct $S_{sym} = D^{-\frac{1}{2}} W D^{-\frac{1}{2}}$ where $D = diag(d_1, d_2, \ldots, d_n)$ and $d_i = \sum_j W_{ij}$
2. Iterate until convergence
   $F^{(t+1)} = \alpha S_{sym} F^{(t)} + (1-\alpha)Y$, where α is an arbitrary parameter belongs to [0,1]
3. Let $F^*$ be the limit of the sequence $\{F^{(t)}\}$.

Next, we look for the closed-form solution of the symmetric normalized graph Laplacian based semi-supervised ranking algorithm. In the other words, we need to show that

$$F^* = \lim_{t \to \infty} F^{(t)} = (1-\alpha)(I - \alpha S_{sym})^{-1} Y$$

Suppose $F^{(0)} = Y$, then

$$F^{(1)} = \alpha S_{sym} F^{(0)} + (1-\alpha)Y$$

$$= \alpha S_{sym} Y + (1-\alpha)Y$$

$$F^{(2)} = \alpha S_{sym} F^{(1)} + (1-\alpha)Y$$

$$= \alpha^2 S_{sym}^2 Y + (1-\alpha)\alpha S_{sym} Y + (1-\alpha)Y$$

$$F^{(3)} = \alpha S_{sym} F^{(2)} + (1-\alpha)Y$$

$$= \alpha^3 S_{sym}^3 Y + (1-\alpha)\alpha^2 S_{sym}^2 Y + (1-\alpha)\alpha S_{sym} Y + (1-\alpha)Y$$

...

Thus, by induction,

$$F^{(t)} = \alpha^t S_{sym}^t Y + (1-\alpha) \sum_{i=0}^{t-1} (\alpha S_{sym})^i Y$$

Since $D^{-\frac{1}{2}} W D^{-\frac{1}{2}}$ is similar to $D^{-1} W$ which is a stochastic matrix, eigenvalues of $D^{-\frac{1}{2}} W D^{-\frac{1}{2}}$ belong to [-1,1]. Moreover, since $0 \leq \alpha \leq 1$, thus

$$\lim_{t \to \infty} \alpha^t S_{sym}^t = 0$$

$$\lim_{t \to \infty} \sum_{i=0}^{t-1} (\alpha S_{sym})^i = (I - \alpha S_{sym})^{-1}$$

Therefore,

$$F^* = \lim_{t \to \infty} F^{(t)} = (1-\alpha)(I - \alpha S_{sym})^{-1} Y$$

Now, from the above formula, we can compute $F^*$ directly.

**Regularization Framework**

In this section, we will develop the regularization framework for the symmetric normalized graph Laplacian based semi-supervised ranking iterative version. First, let's consider the error function

$$E(F) = \left\{ \sum_{i,j=1}^{n} W_{ij} \left\| \frac{F_i}{\sqrt{d_i}} - \frac{F_j}{\sqrt{d_j}} \right\|^2 \right\} + \gamma \sum_{i=1}^{n} \|F_i - Y_i\|^2$$

In this error function $E(F)$, $F_i$ and $Y_i$ belong to $R^1$. Please note that $d_i = \sum_j W_{ij}$, and $\gamma$ is the positive regularization parameter. Hence

$$F = \begin{bmatrix} F_1^T \\ \vdots \\ F_n^T \end{bmatrix} \text{ and } Y = \begin{bmatrix} Y_1^T \\ \vdots \\ Y_n^T \end{bmatrix}$$

Here $E(F)$ stands for the sum of the square loss between the estimated ranking matrix and the initial ranking matrix and the smoothness constraint.

Hence, we can rewrite $E(F)$ as follows

$$E(F) = trace(F^T(I - S_{sym})F) + \gamma trace((F - Y)^T(F - Y))$$

Our objective is to minimize this error function. In the other words, we solve

$$\frac{\partial E}{\partial F} = 0$$

This will lead to

$$(I - S_{sym})F + \gamma(F - Y) = 0$$

$$F - S_{sym}F + \gamma F = \gamma Y$$

$$F - \frac{1}{1+\gamma} S_{sym} F = \frac{\gamma}{1+\gamma} Y$$

$$\left(I - \frac{1}{1+\gamma} S_{sym}\right) F = \frac{\gamma}{1+\gamma} Y$$

Let $\alpha = \frac{1}{1+\gamma}$. Hence the solution $F^*$ of the above equations is

$$F^* = (1 - \alpha)(I - \alpha S_{sym})^{-1} Y$$

Also, please note that $S_{rw} = D^{-1}W$ is not the symmetric matrix, thus we cannot develop the regularization framework for the random walk graph Laplacian based semi-supervised ranking iterative version.

Next, we will develop the regularization framework for the un-normalized graph Laplacian based semi-supervised ranking algorithm. First, let's consider the error function

$$E(F) = \left\{ \sum_{i,j=1}^{n} W_{ij} \|F_i - F_j\|^2 \right\} + \gamma \sum_{i=1}^{n} \|F_i - Y_i\|^2$$

In this error function $E(F)$, $F_i$ and $Y_i$ belong to $R^1$. Please note that $\gamma$ is the positive regularization parameter. Hence

$$F = \begin{bmatrix} F_1^T \\ \vdots \\ F_n^T \end{bmatrix} \text{ and } Y = \begin{bmatrix} Y_1^T \\ \vdots \\ Y_n^T \end{bmatrix}$$

Here $E(F)$ stands for the sum of the square loss between the estimated ranking matrix and the initial ranking matrix and the smoothness constraint.

Hence, we can rewrite $E(F)$ as follows

$$E(F) = trace(F^T L F) + \gamma trace((F-Y)^T(F-Y))$$

Please note that un-normalized Laplacian matrix of the network is $L = D - W$. Our objective is to minimize this error function. In the other words, we solve

$$\frac{\partial E}{\partial F} = 0$$

This will lead to

$$LF + \gamma(F - Y) = 0$$

$$(L + \gamma I)F = \gamma Y$$

Hence the solution $F^*$ of the above equations is

$$F^* = \gamma(L + \gamma I)^{-1} Y$$

**Experimental Results**

## Datasets

In this paper, we use the transaction dataset available from [11]. This dataset contains 390 transactions. Each transaction has 5 features. In the other words, we are given transaction data matrix ($R^{390*5}$) and the annotation (i.e. the ranking) matrix ($R^{390*1}$).

Then we construct the similarity graph from the transaction data. The similarity graph used in this paper is the k-nearest neighbor graph: Transaction $i$ is connected with transaction $j$ if transaction $i$ is among the k-nearest neighbor of transaction $j$ or transaction $j$ is among the k-nearest neighbor of transaction $i$.

In this paper, the similarity function is the Gaussian similarity function

$$s(T(i,:), T(j,:)) = \exp(-\frac{d(T(i,:), T(j,:))}{t})$$

In this paper, $t$ is set to 1 and the 5-nearest neighbor graph is used to construct the similarity graph from the transaction data.

In order to evaluate the performance of the three proposed graph Laplacian based semi-supervised ranking algorithms, we use the default seed set of three transactions (involved in pump-and-dump behaviors) available from [11]. The IDs of these three transactions are 5, 53, and 369.

## Experimental results

In this section, we experiment with the above proposed un-normalized, random walk, and symmetric normalized graph Laplacian ranking methods in terms of accuracy performance measure.

The accuracy performance measure $Q$ is given as follows

$$Q = \frac{True\ Positive + True\ Negative}{True\ Positive + True\ Negative + False\ Positive + False\ Negative}$$

True Positive (TP), True Negative (TN), False Positive (FP), and False Negative (FN) are defined in the following table 1

**Table 1:** Definitions of TP, TN, FP, and FN

|  |  | Predicted Label | |
|---|---|---|---|
|  |  | Positive | Negative |
| Known Label | Positive | True Positive (TP) | False Negative (FN) |
|  | Negative | False Positive (FP) | True Negative (TN) |

All experiments were implemented in Matlab 6.5 on virtual machine. The leave-one-out testing strategy is used to compute the accuracy performance measures of all methods used in this paper. For the default seed set, one transaction is left out and the remaining transactions are used as the core set in the membership queries. Effective ranking methods should report the left-out transaction in top k ranks. The parameter $\mu$ is set to 1 and the parameter $\alpha$ is set to 0.85. The accuracy performance measures of the above proposed methods are given in the following table 2.

**Table 2. The comparison of accuracies of proposed methods**

| Top *k* ranks | | k=10 | k=20 | k = 30 | k=190 |
|---|---|---|---|---|---|
| Accuracy performance measures (%) | Un-normalized | 0 | **33.33** | **66.67** | 66.67 |
| | Random Walk | 0 | 0 | **66.67** | 66.67 |
| | Symmetric Normalized | 0 | **33.33** | **66.67** | **100** |

The results from the above table shows that the symmetric normalized and the un-normalized graph Laplacian based semi-supervised ranking methods outperform the random walk Laplacian based semi-supervised ranking method.

Last but not least, the symmetric normalized graph Laplacian based semi-supervised ranking algorithm can be considered the current state of the art network based ranking method for this irregular trade behaviors detection problem since it achieves the highest accuracy performance measures.

**Conclusions**

We have developed the detailed un-normalized, random walk, and symmetric normalized graph Laplacian based semi-supervised ranking methods applying to irregular trade behaviors detection problem. Experiments show that the un-normalized and the symmetric normalized graph Laplacian based semi-supervised ranking methods are better than the random walk Laplacian based semi-supervised ranking method. Moreover, the symmetric normalized graph Laplacian based semi-supervised ranking method achieves the highest accuracy performance measures in this irregular trade behaviors detection problem.

Recently, to the best of my knowledge, the symmetric normalized graph p-Laplacian based semi-supervised ranking methods have not yet been developed and applied to any practical problems. In the future, we will develop the symmetric normalized graph p-Laplacian based semi-supervised ranking methods and apply these methods to the irregular trade behaviors detection problem.

Moreover, the symmetric normalized graph p-Laplacian based semi-supervised ranking methods can not only be used in irregular trade behaviors detection problem but also in biomarker discovery problem in cancer classification. In specific, given a set of genes (i.e. the queries) involved in a specific disease (for e.g. leukemia), these methods can also be used to find more genes involved in the same disease by ranking genes in gene co-expression network (derived from gene expression data) or the protein-protein interaction network or the integrated network of them. The way that construct the integrated network from gene co-expression network and protein-protein interaction network will be discuss in the future paper. The genes with the highest rank then will be selected and then checked by biologist experts to see if the extended genes in fact are involved in the same disease. This problem is called biomarker discovery problem in cancer classification.

Finally, we know that the un-normalized, symmetric normalized, and random walk graph Laplacian based semi-supervised ranking methods are developed based on the assumption that the rank values of two adjacent nodes in the network are likely to be the same. Hence this assumption can be interpreted as pair of nodes showing the similar patterns and thus sharing edge in the network tends to have same rank values. However, assuming the pairwise relationship between nodes is not complete, the information a group of nodes that shows the very similar patterns and tends to have same rank values is missed. The natural way overcoming the information loss of the above assumption is to represent the data as the hypergraph [13,14]. A hypergraph is a graph in which an edge (i.e. a hyper-edge) can connect more than two vertices.

In [13,14], the symmetric normalized hypergraph Laplacian based semi-supervised ranking method have been developed and successfully applied to text categorization and letter recognition applications. To the best of my knowledge, the hypergraph Laplacian based semi-supervised ranking methods have not yet been applied to irregular trade behaviors detection problem. In the future, we will develop the symmetric normalized, random walk, and un-normalized hypergraph Laplacian based semi-supervised ranking methods and apply these three methods to the irregular trade behaviors detection problem. In the other words, the hypergraph is constructed by applying k-mean clustering method to the transaction data matrix.

**Acknowledgment**

This research is funded by Vietnam National University Ho Chi Minh City (VNU-HCM) under grant number **C2018-42-02.**

**References**

[1] Leangarun, Teema, Poj Tangamchit, and Suttipong Thajchayapong. "Stock price manipulation detection based on mathematical models." *International Journal of Trade, Economics and Finance* 7.3 (2016): 81.

[2] Öğüt, Hulisi, M. Mete Doğanay, and Ramazan Aktaş. "Detecting stock-price manipulation in an emerging market: The case of Turkey." *Expert Systems with Applications* 36.9 (2009): 11944-11949.

[3] Chen, Jing, Bruce J. Aronow, and Anil G. Jegga. "Disease candidate gene identification and prioritization using protein interaction networks." *BMC bioinformatics* 10.1 (2009): 73.

[4] Köhler, Sebastian, et al. "Walking the interactome for prioritization of candidate disease genes." *The American Journal of Human Genetics* 82.4 (2008): 949-958.

[5] Zhou, Denny, et al. "Ranking on data manifolds." *Advances in neural information processing systems*. 2004.

[6] Zhou, Denny, et al. "Learning with local and global consistency." *Advances in neural information processing systems*. 2004.

[7] Brin, Sergey, and Lawrence Page. "The anatomy of a large-scale hypertextual web search engine." *Computer networks and ISDN systems* 30.1-7 (1998): 107-117.

[8] Tran, Loc. "Application of three graph Laplacian based semi-supervised learning methods to protein function prediction problem." *arXiv preprint arXiv:1211.4289* (2012).

[9] Tran, Loc. "The un-normalized graph p-Laplacian based semi-supervised learning method and protein function prediction problem." *Knowledge and Systems Engineering*. Springer, Cham, 2014. 23-35.

[10] Tsuda, Koji, Hyunjung Shin, and Bernhard Schölkopf. "Fast protein classification with multiple networks." *Bioinformatics* 21.suppl_2 (2005): ii59-ii65.

[11] http://www.lobster.wiwi.hu-berlin.de/

[12] Zhu, Xiaojin, and Zoubin Ghahramani. "Learning from labeled and unlabeled data with label propagation." (2002): 1.

[13] Zhou, Dengyong, Jiayuan Huang, and Bernhard Scholkopf. "Beyond pairwise classification and clustering using hypergraphs." *Proceedings of the Neural Information Processing Systems*. 2005.

[14] Zhou, Denny, Jiayuan Huang, and Bernhard Schölkopf. "Learning with hypergraphs: Clustering, classification, and embedding." *Advances in neural information processing systems*. 2007.